%% file: worsharing.tex
\documentclass[conference]{vldb}
\usepackage[color=yellow,obeyFinal]{todonotes}
\usepackage{graphicx}
\usepackage{epstopdf}
\usepackage{float}
\usepackage{varioref}
\usepackage{mathtools}
\usepackage{xspace}
\usepackage{subfigure}
\usepackage{hyperref}
\usepackage{balance}
\usepackage{algorithm}
\usepackage{algpseudocode}
\usepackage{paralist}
\usepackage{cite}
\usepackage[T1]{fontenc}
\usepackage{beramono}
\usepackage{listings}
\usepackage{xcolor}
\usepackage{amsmath}
\usepackage{amsfonts}
\usepackage{amssymb}

\algnewcommand\algorithmicforeach{\textbf{foreach}}
\algdef{S}[FOR]{ForEach}[1]{\algorithmicforeach\ #1\ \algorithmicdo}

\usepackage{amsmath}

\definecolor{dkgreen}{rgb}{0,0.6,0}
\definecolor{gray}{rgb}{0.5,0.5,0.5}
\definecolor{mauve}{rgb}{0.58,0,0.82}

\lstdefinestyle{myScalastyle}{
  frame=tb,
  language=scala,
  aboveskip=3mm,
  belowskip=3mm,
  showstringspaces=false,
  columns=flexible,
  basicstyle={\small\ttfamily},
  numbers=none,
  numberstyle=\tiny\color{gray},
  keywordstyle=\color{blue},
  commentstyle=\color{dkgreen},
  stringstyle=\color{mauve},
  frame=single,
  breaklines=true,
  breakatwhitespace=true,
  tabsize=3,
}

\newcommand{\se}{\text{SE}}
\newcommand{\ce}{\text{CE}}
\newtheorem{definition}{Definition}

\newcommand{\eg}{\textit{e.g.}\xspace}
\newcommand{\ie}{\textit{i.e.}\xspace}

\makeatletter
\def\BState{\State\hskip-\ALG@thistlm}
\makeatother

\usepackage{caption}

\begin{document}

\title{Cache-based Multi-query Optimization \\ for Data-intensive Scalable Computing Frameworks}

\numberofauthors{3}
\author{
\alignauthor
Pietro Michiardi\\
       \affaddr{EURECOM}\\
       \email{michiard@eurecom.fr}
\alignauthor
Damiano Carra\\
       \affaddr{University of Verona}\\
       \email{damiano.carra@univr.it}
\alignauthor
Sara Migliorini\\
       \affaddr{University of Verona}\\
       \email{sara.migliorini@univr.it}
}
\maketitle

\begin{abstract}
In modern large-scale distributed systems, analytics jobs submitted by various users often share similar work, for example scanning and processing the same subset of data. Instead of optimizing jobs independently, which may result in redundant and wasteful processing, multi-query optimization techniques can be employed to save a considerable amount of cluster resources. In this work, we introduce a novel method combining in-memory cache primitives and multi-query optimization, to improve the efficiency of data-intensive, scalable computing frameworks. By careful selection and exploitation of common (sub)expressions, while satisfying memory constraints, our method transforms a batch of queries into a new, more efficient one which avoids unnecessary recomputations. To find feasible and efficient execution plans, our method uses a cost-based optimization formulation akin to the multiple-choice knapsack problem. Extensive experiments on a prototype implementation of our system show significant benefits of worksharing for both TPC-DS workloads and detailed micro-benchmarks.
\end{abstract}

\section{Introduction}
\label{sec:introduction}
\input{01_introduction}

\section{Related Work}
\label{sec:related}
\input{02_related}


\section{Problem Statement}
\label{sec:problem}
\input{03_problem}


\section{Cache-based work sharing}
\label{sec:caching}
\input{04_caching}

\section{Implementation details}
\label{sec:implementation}
\input{05_implementation}


\section{Experimental Evaluation}
\label{sec:evaluation}
\input{06_evaluation}

\section{Conclusion}
\label{sec:conclusion}
\input{07_conclusion}

\section{Acknowledgments}
The authors would like to thank Khoa Nguyen Trong, Duy-Hung Phan and Quang-Nhat Hoang-Xuan
for the valuable discussions and contribution in the early stage of this work.

\clearpage
\bibliographystyle{abbrv}
\bibliography{worsharing}


\end{document}

%% file: 01_introduction.tex
Modern technologies to analyze large amounts of data have flourished in the past decade, starting with general-purpose cluster processing frameworks such as MapReduce \cite{dean2008mapreduce}, Dryad \cite{isard2007dryad} and Spark \cite{zaharia2012resilient}. More recently, a lot of effort has been put in raising the level of abstraction, and allow users to interact with such systems with a relational API, in addition to a procedural one. SQL-like querying capabilities are not only interesting to users for their simplicity, but also bring additional benefits from a wide range of automatic query optimizations, aiming at efficiency and performance.

Currently, such large-scale analytics systems are deployed in shared environments, whereby multiple users submit queries concurrently. In this context, concurrent queries often perform similar work, such as scanning and processing the same set of input data. The research in \cite{gunda2010nectar} on 25 production clusters, estimated that over 35,000 hours of redundant computation could be eliminated per day by simply reusing intermediate query results (approximately equivalent to shutting off 1500 machines daily). It is thus truly desirable to study query optimization techniques that go beyond optimizing the performance of a single query, but instead consider multiple queries, for a more efficient resource utilization, and better aggregate performance.

\emph{Multi-query optimization} (MQO) amounts to find similarities among a set of queries and uses a variety of techniques to avoid redundant work during query execution. For traditional database systems, MQO trades some small optimization overheads for increased query performance, using techniques such as sharing sub-expressions \cite{zhou2007efficient, sellis1988mqo, roy2000efficient}, materialized views selection \cite{goldstein2001optimizing, mistry2001materialized}, and pipelining \cite{dalvi2001pipelining}. Recently, work sharing optimizations operating at query runtime, for staged databases, have also been extensively studied \cite{psaroudakis2013sharing, harizopoulos2005, arumugam2010datapath, giannikis2012shareddb}. The idea of reusing intermediate data across queries or jobs running in a distributed environment has also received significant attention: for MapReduce \cite{mrshare, mqo}, for SCOPE operating on top of Cosmos \cite{silva2012exploiting} and for Massive Parallel Processing (MPP) frameworks \cite{el2015optimization}.

In this paper, we study MQO in the context of distributed computing engines such as Apache Spark \cite{zaharia2012resilient}, with analytics jobs written in SparkSQL \cite{sparksql}, in which relational operators are mapped to stages of computation and I/O. Following the tradition of RDBMSes, queries are first represented as (optimized) logical plans, which are transformed into (optimized) physical plans, and finally run as execution plans. Additionally, modern parallel processing systems, such as Spark, include an operator to materialize in RAM the content of a (distributed) relation, which we use extensively.
Our approach to MQO is that of traditional database systems, as it operates on a batch of queries. However, unlike traditional approaches, it blends pipelining and global query planning with shared operators, using in-memory caching to support worksharing. Our problem formulation amounts to a \emph{cache admission problem}, which we cast as a cost-based, constrained combinatorial optimization task, setting it apart from previous works in the literature. 

We present the design of a MQO component that, given a set of concurrent queries, proceeds as follows. First, it analyzes query plans to find sharing opportunities, using logical plan fingerprinting and an efficient lookup procedure. Then it builds multiple \emph{sharing plans}, using shared relational operators and scans, which subsume common work across the given query set. Sharing plans materialize their output relation in RAM. A cost-based optimization selects best sharing plans with dynamic programming, using cardinality estimation and a knapsack formulation of the problem, that takes into account a memory budget given to the MQO problem. The final step is a global query plan rewrite, including sharing plans which pipeline their output to modified consumer queries of the original input set.

We present a prototype of our system built for SparkSQL, and validate it through a series of experiments. First, using the standard TPC-DS benchmark, we provide an overview of query runtime distributions across a variety of different queries which are optimized using our method: overall, our method achieves up to 80\% reduction in query runtime, when compared to a setup with no worksharing. Then, we proceed with a synthetic evaluation of individual operators, to clarify which ones benefit most from our technique, including when data is materialized on disk according to different formats.
Our main contributions are as follows:
\begin{enumerate}
	\item We propose a general approach to MQO for distributed computing frameworks that support a relational API. Our approach produces \emph{sharing plans}, that are materialized in RAM, aiming at eliminating redundant work and I/O in a given set of queries.
	\item We cast the optimization problem of selecting the best sharing plans as a Multiple-choice Knapsack problem, and solve it efficiently through dynamic programming.
	\item Our ideas materialize into a system prototype, which extends the SparkSQL Catalyst optimizer, and that we evaluated extensively, using macro and micro benchmarks. Our results indicate tangible improvements in terms of aggregate query execution times, while fulfilling the memory budget given to the MQO problem.
\end{enumerate}

The rest of the paper is structured as follows. Section \ref{sec:related} covers related work on multi-query optimization. We introduce and formalize our optimization problem in Section \ref{sec:problem}, and present our methodology in Section \ref{sec:caching}. In Section \ref{sec:implementation}, we provide the implementation details of our prototype based on SparkSQL. We evaluate the performance of our approach in Section \ref{sec:evaluation}. Finally, we conclude the paper in Section \ref{sec:conclusion}.

%% file: 02_related.tex
We now review previous work on MQO, both for traditional RDBMSes and for distributed computing frameworks.

\textbf{MQO in RDBMSes.} Multi-query optimization has been studied extensively \cite{finkelstein1982common, sellis1988mqo, shim1999dynamic, roy2000efficient, dalvi2001pipelining}. More recently, similar subexpressions sharing has been revisited by Zhou et al. in \cite{zhou2007efficient}, who show that reusable common subexpressions can improve query performance. Their approach avoids some limitations of earlier work \cite{finkelstein1982common, roy2000efficient} by (i) considering all generated plans as sharing opportunities to avoid leading to suboptimal plans and (ii) also considering multiple competing covering expressions.
More recently, work sharing at the level of the execution engine has been extensively studied \cite{psaroudakis2013sharing, harizopoulos2005, arumugam2010datapath, giannikis2012shareddb}. The MQO problem is considered at query runtime, and requires a staged database system. Techniques such as pipelining \cite{harizopoulos2005} and multiple query plans \cite{candea2011predictable, candea2009scalable} have proven extremely beneficial for OLAP workloads.

Our work is rooted on such previous literature, albeit the peculiarities of the distributed execution engine (which we also take into account in our cost model), and the availability of an efficient mechanism for distributed caching steer our problem statement apart from the typical optimization objectives and constraints from the literature.

Materialized views can be used in conjunction with MQO to reduce query response times \cite{yang1997algorithms, goldstein2001optimizing, mistry2001materialized}, whereby the contents of materialized views are precomputed and stored to provide alternative, faster ways of computing a query. Nevertheless, due to the nature of RDBMSes, base relations change frequently, leading to view inconsistency and view maintenance costs. For this reason, a broad range of works addressed the problem of materialized view selection and maintenance, including both deterministic \cite{agrawal2000automated, baril2003selection, roy2000efficient} and randomized \cite{zhang1999genetic, derakhshan2006simulated, kalnis2002view} strategies.

In this paper, we focus on analytics queries for systems in which data can be assumed to be static. Such systems are built to support workloads consisting mostly of ad-hoc, long running, scan-heavy queries over data that is periodically loaded in a distributed file system. As such, problems related to view maintenance do not manifest in our setup. Moreover, while materialized views are generally stored permanently (to disk), our approach considers storing intermediate relations in RAM.

\textbf{MQO in Cloud and Massively Parallel Processing (MPP).} Building upon MQO techniques in RDBMSes, Silva et al. \cite{silva2012exploiting} proposed an extension to the SCOPE query optimizer which optimizes cloud scripts containing common expressions while reconciling physical requirements.

In the context of MPP databases, the work in \cite{el2015optimization} presents a comprehensive framework for the optimization of Common Table Expressions (CTEs) implemented for Orca. Compared to our method, we consider not only CTEs but also similar subexpressions to augment sharing opportunities.

\textbf{MQO in MapReduce.} The idea of avoiding redundant processing by batching concurrent MapReduce jobs and make them share some intermediate results was widely studied in \cite{mrshare, mqo, agrawal2008scheduling, bhatotia2011incoop, elghandour2012restore, silva2012exploiting, li2011platform, li2012scalla}. The common denominator of such previous work is that they operate at a lower level of abstraction than we currently do in this paper: they analyze low-level programs that use the procedural API to describe an analytical job. For instance, MRShare \cite{mrshare}, is a sharing framework for MapReduce that first identifies different jobs sharing portions of identical work. These jobs are then transformed into a compound job such that \emph{scan sharing}, \emph{Map Output sharing} and \emph{Map Functions Sharing} can be achieved.

\textbf{Caching to recycle work.} Finally, we consider previous works \cite{azim2017recache, dursun2017revisiting, floratou2016adaptive, ivanova2010architecture, nagel2013recycling} that address the problem of reusing intermediate query results, which is cast as a general caching problem. Our work substantially differs from those approaches in that they mainly focus on cache \emph{eviction}, where past queries are used to decide what to keep in memory, in an on-line fashion. Instead, in this work we focus on the off-line constrained optimization problem of cache \emph{admission}: the goal is to decide the best content to store in the cache, rather than selecting which to evict if space is needed. The only work that considers the reuse of intermediate results when analyzing the overall execution plan of \emph{multiple} queries is \cite{dursun2017revisiting}. Nevertheless, they focus on small problem instances which do not require the general, cost-based approach we present in this work.

\begin{figure*}[ht]
	\centering
	\includegraphics[width=0.92\linewidth]{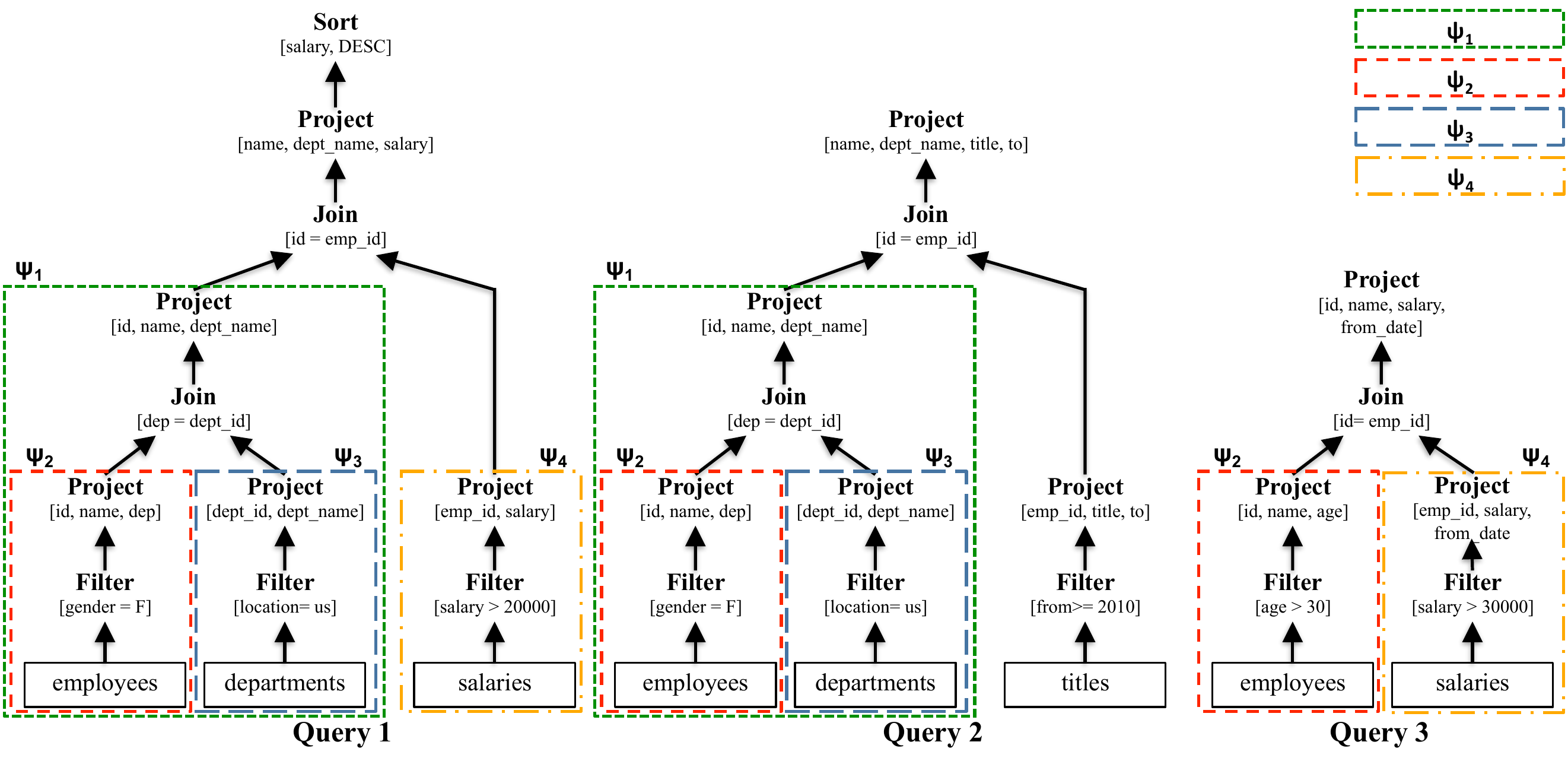}
	\caption{Logical plans for the queries in our running example. Each operator tree has been optimized for query individually. Similar subexpressions (\se{}) inside logical plans are emphasized by dashed boxes surrounding the corresponding sub-tree of each query logical plan. Boxes with the same border color denotes the same SE.}
	\label{fig:common_sub}
\end{figure*}

%% file: 03_problem.tex
This section frames our problem statement and defines a running example that we use in the following sections.

In this paper we focus on the MQO problem only; we gloss over systems aspects that a full-fledged solution should consider as well. In particular, the assumption that a set of concurrent queries is given as an input to the MQO problem hides some complexity and requires careful engineering. This amounts to the design and management of a queue of pending queries, to determine the queue size, and when (given a sufficient number of queued queries) to trigger the MQO. In addition, given a batch of concurrent queries, in case of low concurrency and sufficient available resources, the system should be able to discern whether to execute such queries in parallel, as in a traditional query-centric model, or to apply MQO, as done for example in \cite{psaroudakis2013sharing}. Finally, given an optimized query set, produced by our MQO strategy, the system should determine in which order to schedule the new batch of queries, given a performance objective.

We now introduce a simple running example, that is rich enough to illustrate the MQO problem. Consider the following three concurrent queries:

\begingroup
\fontsize{7pt}{8pt}
\selectfont
\begin{verbatim}
QUERY 1:
SELECT name, dept_name, salary
FROM employees, departments, salaries
WHERE dep = dept_id
    AND id = emp_id
    AND gender = 'F'
    AND location = 'us'
    AND salary > 20000
ORDER BY salary DESC

QUERY 2:
SELECT name, dept_name, title,
    to as title_expired_on
FROM departments, employees, titles
WHERE dep = dept_id
    AND id = emp_id
    AND gender = 'F'
    AND location = 'us'
    AND from >= 2010

QUERY 3:
SELECT id, name, salary, from_date
FROM employees, salaries
WHERE id = emp_id
    AND age > 30
    AND SALARY > 30000
\end{verbatim}
\endgroup

We use Figure\,\ref{fig:common_sub} to illustrate the optimized operator trees (logical plans) of the queries in the above example. The leaf nodes represent the base relations. Each intermediate node is a relational algebra operator (\emph{Selection, Projection, Join}, etc.). The arrows between nodes indicate data flow. Our MQO strategy uses such optimized logical plans to produce new plans -- whose aim is to exploit sharing opportunities by caching in RAM distributed relations -- which are then translated into physical plans for execution.

First, we see that the three queries can share the scan of the \textbf{employees}, \textbf{departments} and \textbf{salaries} relations. Hence, a simple approach to work sharing would be to inject a cache operator in Query\,1, which would steer the system to serve input relations from RAM instead of reading them from disk, when executing Query\,2 and 3.
A more refined approach could be to find common work (not only common I/O), in the form of \emph{similar subexpressions} (\se{}) among the queries from the example, such as filtering and projecting records, joining tables, etc, and materialize intermediate results in RAM, to speed-up query runtime by re-using such intermediate relations.

Figure\,\ref{fig:common_sub} illustrates four examples of similar \se{}s, which are labelled as $\psi_i, i = 1,2,3,4$ (we explain the meaning of this label in the next section). For example, consider the subexpression labelled as $\psi_2$: all three queries share the same sub-tree structure, in the form \emph{Project}$_p$(\emph{Filter}$_f$(\textbf{employees})), but use different filtering predicates and projections. In principle, it is thus possible to save reading, parsing, filtering and projecting costs on the \textbf{employees} relation: by caching the intermediate output of a general form of subexpression, which subsumes the three similar sub-trees in each query. Such costs would be payed only once, and the cached intermediate relation could serve three \emph{consumer} queries. To achieve that, we need to build a \emph{covering expression} (\ce{}) that combines the different variants of the predicates appearing in the operators, for instance considering $\psi_2$ the corresponding \ce{} could be:
\begin{center}
\emph{Project}$_{\text{id, name, dep, age}}$(\emph{Filter}$_{\text{gender=F}\,\lor\, \text{age>30}}$(\textbf{employees}))
\end{center}

In a similar vein, the \se{}s labelled as $\psi_3$ and $\psi_4$ share the projection and filtering on \textbf{department} and \textbf{salaries} relations, respectively.

We anticipate that, in the context of our work, it is possible to rank some \se{}s according to the benefits they bring, in terms of reducing redundant work. For instance, the \se{} \emph{Project}$_p$(\emph{Filter}$_f$(\textbf{employees})) leads to additional savings when compared to the \se{} \emph{Filter}$_f$(\textbf{employees}), and caching the intermediate relation of the corresponding \ce{} results in a smaller memory footprint because of its selectivity.
More to the point, we now consider the \se{} labelled as $\psi_1$ in Figure\,\ref{fig:common_sub}: Query\,1 and 2 share a common sub-tree in their respective logical plans, that involves selections, projections and joins. In this case, selecting this \se{} as a candidate to build a \ce{} between Query\,1 and 2 contributes to decreased scanning, computation and \emph{communication costs}. However, since caching a relation in RAM bears its own costs and must satisfy capacity constraints, materializing in RAM the output of the \ce{} might reveal not beneficial after all. For example, a join operator could potentially produce an intermediate relation too big to fit in RAM.

Overall, given an input query set, our problem amounts to explore a potentially very large search space, to identify \se{}s, to build the corresponding \ce{}s -- which we also call \emph{sharing plans}, and to decide which \ce{}s to include in the optimized output plan. Our MQO strategy aims at reducing the search space to build \ce{}s by appropriately pruning \se{}s according to their rank. Furthermore, a cost-based selection of candidate \ce{}s must ensure memory budget constraints to be met. In the following Section, we delve into the detail of our MQO approach.

%% file: 04_caching.tex
This section describes our approach to MQO, using a caching operator to materialize in RAM intermediate (distributed) relations that belong to a \emph{sharing plan}. We assume a set of concurrent queries submitted by multiple users to be parsed, analyzed and individually optimized by a query optimizer. Our MQO method operates on a set of optimized logical plans corresponding to the set of input queries, that we call the \emph{input set}.

We approach the MQO problem with the following steps:
\begin{enumerate}
	\item \textbf{Similar subexpressions identification.} The goal of this phase is to identify all common and similar subexpressions in the input set, as discussed in Section \ref{sec:common_sub}. In short, we compute an \emph{operator fingerprint} for each operator in the logical plan of every query and store it in a fingerprint table. Two (or more) operators sharing the same fingerprint constitute a \se{}. Identified \se{}s are candidates for building covering expressions (\ce{}s) in the next step.
	\item \textbf{Building Covering subexpressions, \emph{a.k.a.} sharing plans.} Given all \se{}s identified in an input set, the goal of this phase is to construct one or more groups of \ce{}s representing candidate sharing plans, as discussed in Section \ref{sec:covering_subexpression}. Since the search space for building \ce{}s and for their subsequent selection can be very large, in this phase our approach prunes bad \se{}s, in an attempt to produce few, good \ce{}s candidates.
	\item \textbf{Sharing plan selection.} The goal of this phase is to select the best combination of \ce{}s, using estimated costs and memory constraints, as shown in Section \ref{sec:cbo}. The output of this phase is a series of sharing plans which use \emph{shared operators} (covering those of the underlying \se{}s) and materialize their output relation in RAM using a cache operator. For this reason, we sometimes refer to such \ce{}s as \emph{caching plans}. We model this step as a Multiple-Choice Knapsack problem, and use dynamic programming to solve it.
	\item \textbf{Query rewriting.} The last step to achieve MQO is to rewrite the input query set such as to use selected sharing plans, as shown in Section \ref{sec:query_rewriting}. Essentially, cached relations pertaining to a \ce{} are \emph{pipelined} to those queries that can be rewritten using that \ce{}. The output of this phase is a new set of rewritten queries, that subsume the input set, although their execution is not guaranteed to be in the original ordering.
\end{enumerate}

\subsection{Similar Subexpression Identification}
\label{sec:common_sub}
Finding similar subexpressions, given an input set of logical plans, has received considerable attention in the literature. What sets our approach apart from previous works lies behind the very nature of the resource we use to achieve work sharing: memory is limited, and the overall MQO process we present is seen as a constrained optimization problem, which strives to use caching with parsimony.
Thus, we use a general rule of thumb that prefers a large number of \ce{}s (built from the corresponding \se{}s) with small memory footprints instead of a small number of \ce{}s with large memory requirements. This rule of thumb is also in line with low-level systems considerations: data materialization in RAM is \emph{not} cost-free, and current parallel processing frameworks are sometimes fragile, when it comes to memory management under pressure.

Armed with the above considerations, we first consider the input to our task: we search \se{}s given a set of ``locally optimized'' query plans, which are represented in a tree form. Such input plans have been optimized by applying common rules such as early filtering, predicate push-down, plan simplification and collapsing \cite{sparksql}. The natural hierarchy of an optimized logical plan, in general, implies that the higher in the tree an operator is, the less the data flowing from its edges. Hence, similar subexpressions that are found higher in the plan hierarchy are preferred because they potentially exhibit smaller memory footprints, should their output relation be cached.

Additional considerations are in order. Some operators produce output relations that are not easy to materialize in RAM: for example, binary operators such as join, generally produce large outputs that would deplete memory resources if cached. Thus, when searching for \se{}s, we recognize ``cache unfriendly'' operators and preempt them for being considered as valid candidates, either by selecting \se{}s that appear lower in the logical plan hierarchy (e.g., which could imply caching the input relations of a join), or by selecting \se{}s that subsume them (e.g., which could imply caching a relation resulting from filtering a join output). Currently, we treat the join, Cartesian product and union as ``cache unfriendly'' operators. This means that our method does not produce \se{}s \emph{rooted at} cache unfriendly operators; moreover, cache unfriendly operators can be shared inside a common \se{} only when they are syntactically equal.\footnote{Our method can be easily extended for sharing similar join operators, for example by applying the ``equivalence classes'' approach used in \cite{zhou2007efficient}. Despite technical simplicity, our current optimization problem formulation would end-up discarding such potential \se{}s, due to their large memory footprints. Hence, we currently preempt such \se{}s from being considered.}
In the following, we provide the necessary definitions that are then used to describe the identification of \se{}s.

\begin{definition}[Sub-tree]
Given a logical plan of a query $Q$, represented as a tree $\tau^{Q}$ where leaf nodes are base relations and each intermediate node is a relational algebra operator, a sub-tree $\tau_s^{Q}$ of $\tau^{Q}$ is a continuous portion of the logical plan of $Q$ containing an intermediate node of $\tau^{Q}$ and all its descendant in $\tau^{Q}$. In other words, a sub-tree includes all the base relations and operators that are necessary to build its root.
\end{definition}


In the following, if the context is clear, we denote a sub-trees simply as $\tau$, without indicating from which query it has been derived.

Given any two sub-trees, we need to determine if they have the same \emph{structure} in terms of base relations and operators.
To this aim, we define a similarity function based on a modified Merkle Tree (also known as \emph{hash tree})\cite{mtree}, whereby each internal node identifier is the combination of identifiers of its children. More specifically, given an operator $u$, its identifier, denoted by ID($u$), is given by:
\begin{equation*}
\text{ID}(u) =
	\begin{cases}
		(u.label) &\text{$u \in$ \{filter, project,}\\
			      &\text{~~~~~~~~input relation\}}\\
		(u.label, u.attributes) &\text{otherwise}.
	\end{cases}
\end{equation*}
Notice that this definition makes a distinction between loose and strict identifier.
A loose identifier, such that used for projections and selections, allows the construction of a \emph{shared operator} that subsumes the individual attributes with more general ones, which allows sharing computation among \se{}s. Instead, a strict identifier, such that used for all other operators (including joins and unions), imposes strict equality for two sub-graphs to be considered \se{}s. In principle, this restricts the applicability of a shared operator. However, given the above considerations about cache unfriendly operators, our approach still shares both I/O and computation.

\begin{definition}[Fingerprint]
Given a sub-tree $\tau$, its \emph{fingerprint} is computed as
\begin{equation*}
	\mathcal{F}(\tau) =
	\begin{cases}
		h(\mathrm{ID}(\tau_{\mathrm{root}})) &\tau_{\mathrm{root}} = \mathrm{leaf}\\
		h(\mathrm{ID}(\tau_{\mathrm{root}})|\mathcal{F}(\tau_{\mathrm{child}})) &\tau_{\mathrm{root}} = \mathrm{unary}\\
		h(\mathrm{ID}(\tau_{\mathrm{root}})|\mathcal{F}(\tau_{\mathrm{l.child}})|\mathcal{F}(\tau_{\mathrm{r.child}})) &\tau_{\mathrm{root}} = \mathrm{binary}\\
	\end{cases}
\end{equation*}
where $h()$ is a robust cryptographic hash function, and the operation $|$ indicates concatenation.
\end{definition}

The fingerprint $\mathcal{F}(\tau)$ is computed recursively starting from the root of the sub-tree ($\tau_{\mathrm{root}}$), down to the leaves (that is, input relations). If the root is a unary operator, we compute the fingerprint of its child sub-tree ($\tau_{\mathrm{child}}$), conversely in case of a binary operator, we consider the left and right sub-trees ($\tau_{\mathrm{l.child}}$ and $\tau_{\mathrm{r.child}}$). For the sake of an uncluttered notation, we omit an additional sorting which ensures the isomorphic property for binary operators: for example,  \emph{TableA join TableB} and \emph{TableB join TableA} are two isomorphic expressions, and have the same fingerprint.

We are now ready to define what a \emph{similar subexpression} (denoted as $\omega$) is.

\begin{definition}[Similar subexpression]
\label{def.subexpr}
A similar\\
subexpression (\se{}) $\omega$ is a set of sub-trees that have the same fingerprint $\psi$, \ie
$\omega = \{\tau_i \ |\  \mathcal{F}(\tau_i) = \psi\}$.
\end{definition}

Algorithm\,\ref{alg:common_sub_alg} provides a pseudo-code of our procedure to find, given a set of input queries, the \se{}s according to Definition\,\ref{def.subexpr} that will be the input of the next phase, the search for covering expressions. The underlying idea is to avoid a brute-force search of fingerprints, which would produce a large number of \se{}s. Instead, by proceeding in a top-down manner when exploring logical plans, we produce fewer \se{}s candidates, by interrupting the lookup procedure as early and as ``high'' as possible.

\begin{algorithm}[t]
\caption{Similar subexpressions identification}\label{alg:common_sub_alg}
Input: Array of logical plans (trees), threshold $k$\\
Output: Set $S$ of \se{}s $\omega_i$
\begin{algorithmic}[1]
\Procedure{IdentifySEs}{$[\tau^{Q_1}, \tau^{Q_2}, ... \tau^{Q_N}]$}
	\State FT $\leftarrow \emptyset$ \label{FT}
	\ForEach{$\tau \in [\tau^{Q_1}, \tau^{Q_2}, ... \tau^{Q_N}]$}
		\State nodeToVisit $\leftarrow$ \Call{Add}{$\tau$} \label{generator}
		\While{nodeToVisit \text{not empty}}
			\State $\tau^{\text{curr}} \leftarrow$ \Call{Pop}{nodeToVisit}
			\State $\psi \leftarrow \mathcal{F}(\tau^{\text{curr}})$
			\If {\Call{CacheFriendly}{$\tau_{\text{root}}^{\text{curr}}$}} \label{ifcachefriendly}
				\State FT.\Call{AddValueSet}{$\psi$, $\tau^{\text{curr}}$} \label{addvalue}
			\EndIf
			\If {(!\,\Call{CacheFriendly}{$\tau_{\text{root}}^{\text{curr}}$} $\lor$\\
			{~~~~~~~~~~~~~~~~~~~~}\Call{ContainsUnfriendly}{$\tau^{\text{curr}}$})}
				\State nodeToVisit $\leftarrow$ \Call{Add}{$\tau_{\text{children}}^{\text{curr}}$} \label{addchildren}
			\EndIf
		\EndWhile
	\EndFor
	\State $S \leftarrow \emptyset$
	\ForEach{$\psi \in$ FT.\Call{Keys}{}}
		\If {|FT.\Call{GetValue}{$\psi$}| $\geq k$}
			\State $S \leftarrow S \, \cup \,$ FT.\Call{GetValue}{$\psi$}
		\EndIf
	\EndFor
	\State \Return $S$
\EndProcedure
\end{algorithmic}
\end{algorithm}


The procedure uses a fingerprint table FT (line \ref{FT}) to track \se{}s: this is a HashMap, where the key is a fingerprint $\psi$, and the value is a set of subtrees. Each logical plan from the input set of queries is examined in a depth-first manner. We first consider the whole query tree (line \ref{generator}) and check if its root is a cache-friendly operator: in this case, we add the tree to the \se{}s identified by its fingerprint. The method \textsc{AddValueSet($\psi$, $\tau$)} retrieves the value (which is a set) from the HashMap FT given the key $\psi$ (line \ref{addvalue}), and adds the subtree $\tau$ to such a set -- if the key does not exists, it adds it and create a value with a set containing the subtree $\tau$.
If the root is not a cache-friendly operator, or the logical plan contains a cache-unfriendly operator, then we need to explore the subtrees (line \ref{addchildren}), \ie we consider the root's child (if the the operator at the root is unary) or children (otherwise).

At the end, we extract the set of \se{}s from the HashMap FT: we consider the \se{}s bigger than a threshold $k$ (\eg with at least two subtrees from two queries) in order to focus on \se{}s that offer potential work sharing opportunities.

Going back to our running example, Algorithm\,\ref{alg:common_sub_alg} outputs a set of \se{}s as follows $\{ \omega_1, \omega_2, \omega_3, \omega_4 \}$ -- in Fig.\,\ref{fig:common_sub} the sub-trees corresponding to them are labelled $\psi_1, \psi_2, \psi_3$ and $\psi_4$, where $\psi_i$ is the fingerprint of \se{} $ \omega_i$. For instance, $\omega_1$ contains two sub-trees (one from Query\,1, and one from Query\,2), while $\omega_2$ contains three sub-trees, one from each query.

\subsection{Building Sharing Plans}
\label{sec:covering_subexpression}
Given a list of candidate \se{}s, the goal of this phase is to build covering subexpressions (\ce{}s) corresponding to identified \se{}s, and generate a set of candidate groups of \ce{}s for their final selection.


\vspace{1mm}
\noindent
{\bf Covering subexpressions.}
For each similar sub-query in the same \se{} $\omega_i$, the goal is to produce a new plan to ``cover'' all operations of each individual sub-query.

Recall that all sub-trees $\tau_j$ within a \se{} $\omega_i$ share the same sub-query plan fingerprint: that is, they operate on the same input relation(s) and apply the same relational operators to generate intermediate output relations. If the operator attributes are exactly the same across all $\tau_j$, then the \ce{} will be identical to any of the $\tau_j$. In general, however, operators can have different attributes or predicates. In his case, the \ce{} construction is slightly more involved.

\begin{figure}[!t]
	\centering
	\includegraphics[width=1.0\linewidth]{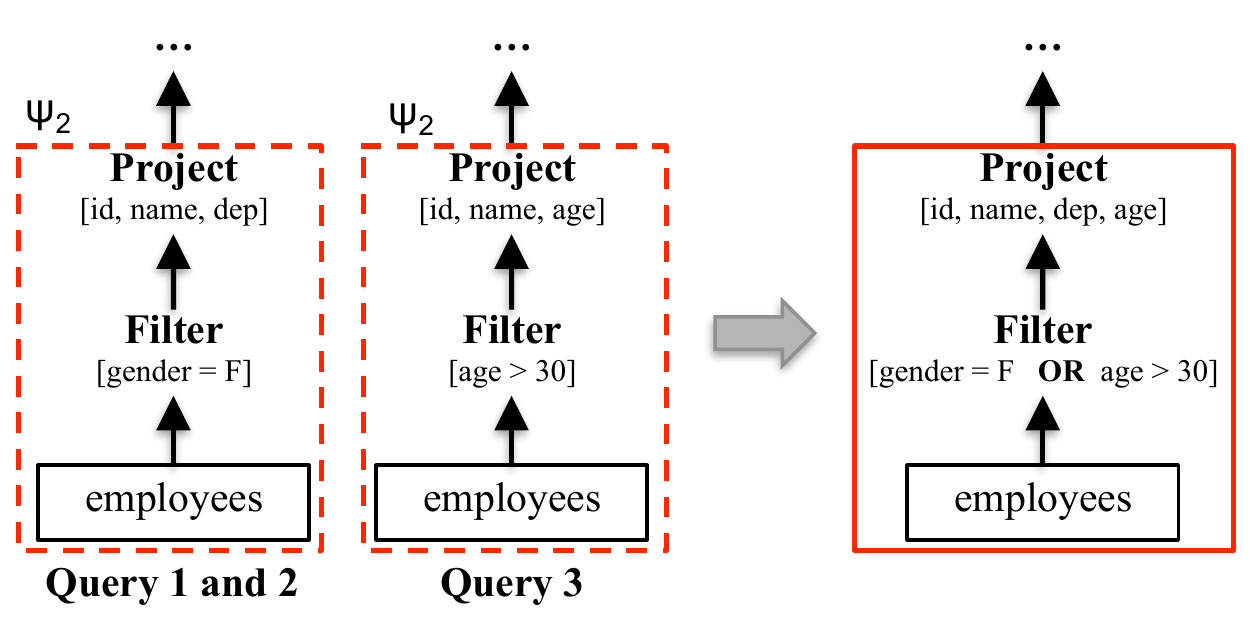}
	\caption{Building covering expression example. The first and second trees are two similar subexpressions. The third tree is the covering subexpression.}
	\label{fig:covering}
\end{figure}

First, we note that, by construction, the only shared operators we consider are projections and selections. Indeed, for \emph{cache unfriendly} operators, the \se{} identification phase omits their fingerprint from the lookup procedure (see Algorithm\,\ref{alg:common_sub_alg}, lines \ref{ifcachefriendly}-\ref{addvalue}). Nevertheless, they could be included within a subtree, but they are in any case ``surrounded'' by cache-friendly operators (see for instance in Fig.\,\ref{fig:common_sub}, the \se{} labeled as $\psi_1$). As a consequence, a \ce{} can be constructed in a top-down manner, by ``OR-ing'' the filtering predicates and by ``unioning'' the projection columns of the corresponding operators in the \se{}. The \ce{} thus produces and materializes all output records that are needed for its consumer queries\footnote{For the sake of readability, we omit the description of several other optimizations -- such as the removal of duplicate predicates -- that we have implemented.}. Fig.\,\ref{fig:covering} illustrates an example of \ce{} for a simple \se{} of two sub-queries taken from the running example shown in Fig.\,\ref{fig:common_sub}. In particular, we consider the \se{} labeled as $\psi_2$.

The resulting \ce{} contains the same operators as the subtrees $\tau_j \in \omega_i$, but with modified predicates or attribute lists.

In general, we can build a \ce{}, which we denote with $\Omega_i$, from a \se{} $\omega_i$, by applying a transformation function $f()$, $[\tau_1, ... \tau_m] \xrightarrow{f()} \tau_i^*$, which trasforms a collection of similar sub-trees to a single, \emph{covering} sub-tree $\tau_i^*$. Note that the resulting covering sub-tree has the fingerprint of the sub-trees in $\omega_i$.

\begin{definition}[Covering subexpression]
\label{def.covexpr}
A Covering
subexpression (\ce{}) $\Omega_i = f(\omega_i)$ is a sub-tree $\tau_i^*$ derived from the \se{} $\omega_i$ by applying the transformation $f()$, with $\mathcal{F}(\tau_i^*) = \mathcal{F}(\tau_j) \,\forall \tau_j \in \omega_i$, such that all $\tau_j \in \omega_i$ can be derived from $\tau_i^*$.
\end{definition}

In summary, the query plan $\tau_i^*$ that composes $\Omega_i$ contains the same nodes as any subtree $\tau_j \in \omega_i$, changing the predicates of the selections (OR of all the predicates in $\tau_j$) and projections (union of all the predicates in $\tau_j$).

Once the set of \ce{}s, $\Omega = \{\Omega_1, \Omega_2, \dots \}$, has been derived from the corresponding set of \se{}s, $\omega = \{\omega_1, \omega_2, \dots \}$, we need to face the problem of \ce{} \emph{selection}. The main question we need to answer is: among the \ce{}s contained in the set $\Omega$, which ones should be cached? Each \ce{} covers different portions of the query logical plans, therefore a \ce{} may include another \ce{}. Looking at the running example shown in Fig.\,\ref{fig:common_sub}, we have that $\Omega_1$ (derived from $\omega_1$, in the figure labeled with $\psi_1$) contains $\Omega_3$ (derived from $\omega_3$ and labeled in the figure with $\psi_3$). If we decide to store $\Omega_1$ in the cache, it becomes questionable to store $\Omega_3$ as well.

The next step of our process is then to identify the potential combinations of \emph{mutually exclusive} \ce{}s that will be the input of the optimization problem: each combination will have a \emph{value} and \emph{weight}, where the value provides a measure of the work sharing opportunities, and the weight indicates the amount of space required to cache the \ce{} in RAM. We start considering how to compute such values and weights, and we proceed with the algorithm to identify the potential combination of \ce{}s.

\vspace{1mm}
\noindent
{\bf \ce{} value and weight: a cost-based model.}
As in traditional database systems, we use cardinality estimation and cost modeling to reason about the benefit of using \ce{}s. The objective is to estimate if a given \ce{}, that could serve multiple consumer queries, yields lower costs than executing \emph{individually} the original queries it subsumes.

The \emph{cardinality estimator} component analyzes relational operators to estimate their output size. To do so, it first produces statistics about input relations and their columns. At relation level, it obtains the number of records and average record size. At column level, it collects the min and max values, approximates column cardinality and produces an equi-width histogram for each column.

The \emph{cost estimator} component uses the results from cardinality estimation to approximate a (sub) query execution cost. We model the total execution cost of a (sub) query as a combination of CPU, disk and network I/O costs. Hence, given a sub-tree $\tau_j$, we denote by $\mathcal{C}_E(\tau_j)$ the execution cost of sub-tree $\tau_j$.
This component recursively analyzes, starting from the root of sub-tree $\tau_j$, relational operators to determine their cost (and their selectivity), which is the multiplication between predefined constants (representative of the compute cluster running the parallel processing framework) and the estimated number of input and output records. Given a \se{} $\omega = \{\tau_1, \tau_2, \dots, \tau_m\}$, the total execution cost $\mathcal{C}(\omega_i)$ related to the execution of all similar sub-trees $\tau_j \in \omega_i$ without the work-sharing optimization is given by
\begin{equation}
\label{eq:se_cost}
	\mathcal{C}(\omega_i) = \sum_{j=1}^{m} \mathcal{C}_E(\tau_j).
\end{equation}
Instead, the cost of using the corresponding \ce{} $\Omega_i$ must account for both the execution cost of the common sub-tree $\tau_i^*$, and materialization ($\mathcal{C}_W$) and retrieving ($\mathcal{C}_R$) costs \emph{associated to the cache operator we use in our approach}, which accounts for write and read operations:
\begin{equation}
\label{eq:ce_cost}
	\mathcal{C}(\Omega_i) = \mathcal{C}_E(\tau_i^*) + \mathcal{C}_W(|\tau_i^*|) + m \cdot \mathcal{C}_R(|\tau_i^*|),
\end{equation}
where both $\mathcal{C}_W(|\tau_i^*|)$ and $\mathcal{C}_R(|\tau_i^*|)$ are functions of the cardinality $|\tau_i^*|$ of the intermediate output relation obtained by executing $\tau_i^*$. Eq.\,\ref{eq:ce_cost} indicates that retrieving costs are ``payed'' by each of the $m$ \emph{consumer} queries from the \se{} $\omega_i$ that can use the corresponding \ce{} $\Omega_i$.\footnote{In light of the end-to-end MQO process, the last phase amounts to rewrite the queries in the input set to use selected \ce{}s. Such rewrite can introduce additional work, which we currently neglect in our modeling approach: indeed, query rewriting involves highly selective operations, with low cost. This means we assume the dominating cost to be that of reading from RAM, which we found experimentally to be true.}

Given the above costs, we can derive the \emph{value} of a \ce{} $\Omega_i$, denoted by $v(\Omega_i)$, as the difference between the cost of an unoptimized set of sub-trees (execution of $\omega_i$) and the cost of the corresponding \ce{} $\Omega_i$:
\begin{equation}
\label{eq:ce_value}
	v(\Omega_i) = \mathcal{C}(\omega_i) - \mathcal{C}(\Omega_i).
\end{equation}
From Equations\,\ref{eq:se_cost} and \ref{eq:ce_cost}, we note that $v(\Omega_i)$ is an increasing function in $m$. Indeed, the more similar sub-queries a \ce{} can serve, the higher its value.

Along with the value, we need to associate to a \ce{} also a \emph{weight}, since the memory is limited and we need to take into account if a \ce{} can fit in the cache. The weight, denoted by $w(\Omega_i)$ is the size  required to cache in RAM the output of $\Omega_i$, \ie $w(\Omega_i) = | \tau_i^* | \overset{\Delta}{=} | \Omega_i |$.

Having defined the \ce{} value and weight, we describe next the algorithm to identify the potential combination of \ce{}.

\vspace{1mm}
\noindent
{\bf Generating the candidate set of \ce{}s.} Next, we focus on the problem of generating a combinatorial set of \ce{}s, with their associated value and weight, to be given as an input to the multi-query optimization solver we have designed. Given the complexity of the optimization task, our goal is to produce a small set of valuable alternative options, which we call the candidate set of \ce{}s. We present an algorithm to produce such a candidate set, but first illustrate the challenges it addresses using the example shown in Figure\,\ref{fig:common_sub}.

Let's focus on \ce{} $\Omega_1$ (corresponding to the sub-trees labeled as $\psi_1$). A naive enumeration of all possible choices of candidate \ce{} to be cached leads to the following, \emph{mutually exclusive} options: (i) $\Omega_1$, (ii) $\Omega_2$, (iii) $\Omega_3$, (iv) both $(\Omega_2$,$\Omega_3)$, (v) both $(\Omega_1$,$\Omega_2)$, and (vi) both $(\Omega_1$,$\Omega_3)$. Intuitively, however, it is easy to discern valuable from wasteful options. For example, the compound \ce{} $(\Omega_1,\Omega_2)$ could be a good choice, since $\Omega_2$ can be cached to serve query 1 and 2 -- and of course used to build $\Omega_1$ -- and for query 3. Conversely, caching the compound $(\Omega_1,\Omega_3)$ brings less value, since it only benefits query 1 and query 2, but costs more than simply caching $\Omega_1$, which also serves both query 1 and 2.

It is thus important to define \emph{how to compute the value and weight of compound} \ce{}. In this work we only consider compound \ce{}s for which value and weight are \emph{additive} in the values and weights of their components. This property is achieved by considering compounds of \emph{disjoint} \ce{}s, \emph{i.e.}, those that have no common sub-trees.

For example, consider the two \ce{}s $\Omega_1$ and $\Omega_2$, and the sub-trees used to build them. The \ce{} $\Omega_2$ is included in $\Omega_1$, but only some of the originating sub-trees of $\Omega_2$ are included in the originating sub-trees of $\Omega_1$ (in particular, the ones in query 1 and 2, but not in query 3). Given our definition of the value and the weight of \ce{}s, the value and the weight of the compound $(\Omega_1,\Omega_2)$ may not be equal to the sums of the values and of the weights of each individual \ce{}, since part of the \ce{} need to be reused to compute different sub-trees. Thus, we discard this option from the candidate set.

\begin{algorithm}[t]
\caption{Algorithm to generate CE candidates.}\label{alg:ce_candidate}
Input: Set $\Omega$ of \ce{}s\\
Output: Set of Knapsack items (potential \ce{}s)
\begin{algorithmic}[1]\raggedright
\Procedure{GenerateKPitems}{$\Omega = \{\Omega_1, \Omega_2, \dots \}$}
	\State $\Omega^{\text{exp}} \leftarrow \emptyset$
	\While{$\Omega$ \text{not empty}}
		\State $\Omega_{i} \leftarrow$  \Call{PopLargest}{$\Omega$} \label{CEoutLargest}
		\State DescSet $\leftarrow$ \Call{FindDescendant}{$\Omega_{i}$, $\Omega$} \label{CEoutDesc}
		\State Group$_i \leftarrow [ \Omega_i ] \, \cup \, $\Call{Expand}{DescSet} \label{CEdisjoint}
		\State $\Omega^{\text{exp}} \leftarrow \Omega^{\text{exp}} \, \cup \, \{$Group$_i\}$ \label{CEoutCE}
		\State \Call{Remove}{DescSet, $\Omega$} \label{CEoutRemove}
	\EndWhile
	\State \Return $\Omega^{\text{exp}}$
\EndProcedure
\end{algorithmic}
\end{algorithm}

Algorithm\,\ref{alg:ce_candidate} generates the candidate input for the optimization solver as a set of non-overlapping groups of \ce{}s; then, the optimization algorithm selects a single candidate for each group in order to determine the best set of \ce{}s to store in memory.
Given the full set of $\Omega$ of \ce{}s as input, we consider \ce{} $\Omega_i$ starting from the root of the logical plan and remove it from the set (line\, \ref{CEoutLargest}). We then look for its descendants from the input set $\Omega$, \ie all the \ce{}s contained in $\Omega_i$ (line\, \ref{CEoutDesc}). With a \ce{} and its descendant, we build a list of options that contains (i) the \ce{} itself and its individual descendants, and (ii) all the compounds of \emph{disjoint} descendant \ce{}s (line\,\ref{CEdisjoint} and \ref{CEoutCE}). We then remove the descendant from $\Omega$ and continue the search for other groups.

Considering our running example, we start from $\Omega = \{\Omega_1, \Omega_2, \Omega_3, \Omega_4\}$. The ``largest'' \ce{} is $\Omega_1$, and its descendants are $\Omega_2$ and $\Omega_3$, therefore the list of mutually exclusive options for this group would be $\left[ \Omega_1, \Omega_2, \Omega_3, (\Omega_2,\Omega_3)\right]$. The final output of Algorithm\,\ref{alg:ce_candidate} then is:
\begin{equation}
\label{listCErunning}
\left\{ \left[ \Omega_1, \Omega_2, \Omega_3, (\Omega_2,\Omega_3)\right], \left[\Omega_4\right]\right\},
\end{equation}
where the notation $(\cdot, \cdot)$ indicates a compound \ce{}, and $[ \cdot, \cdot ]$ indicates a group of related \ce{}s.

Note that a \ce{} may be part of more than one larger \ce{}: to keep the algorithm simple, we consider only the largest ancestor for each \ce{}. To each option, we associate the value and the weight (in case of a compound, the sum of each component), that will be used by the optimization solver.

\vspace{3mm}

\subsection{Sharing Plan Selection}
\label{sec:cbo}
Next, we delve into our MQO problem formulation. In this work, we model the process that selects which sharing plan to use as a Multiple-choice Knapsack problem (MCKP) \cite{sinha1979multiple}. Essentially, the knapsack contains items (that is, sharing plans or CEs) that have a \emph{weight} and a \emph{value}. The knapsack capacity is constrained by a constant $c$: this is representative of the memory constraints given to the work sharing optimizer. Hence, the sum of the weights of all items placed in the knapsack cannot exceed its capacity $c$.

Our problem is thus to select which set of \ce{}s (single, or compound) to include in the knapsack. The output of the previous phase (and in particular, the output of Algorithm\,\ref{alg:ce_candidate}) is a set containing $m$ groups of mutually exclusive options, or items. Each group $G_i$, $i = 1, 2, \dots, g$, contains $|G_i|$ items, which can be single \ce{} or compounds of \ce{}s. For instance, looking at our running example, the output shown in Eq.\,(\ref{listCErunning}) contains $g=2$ groups: the first group has 4 items, the second group just one item. Given a group $i$, each item $j$ has a value $v_{i,j}$ and a weight $w_{i,j}$ computed as described in Sect.\,\ref{sec:covering_subexpression}.

The MCKP solver needs to choose \emph{at most} one item from each group such that the total value is maximized, while the corresponding total weight must not exceed the capacity $c$. More formally, the problem can be cast as following:
\begin{equation}
\label{e:maxproblem}
\begin{aligned}
\text{Maximize~~~} & \sum_{i=1}^{g}\sum_{j=1}^{|G_i|} v_{i,j}x_{i,j}\\
\text{subject to~~~} & \sum_{i=1}^{g}\sum_{j=1}^{|G_i|} w_{i,j}x_{i,j} \leq c\\
 & \sum_{j=1}^{|G_i|}x_{i,j} \leq 1, \forall i = 1\ldots g\\
& x_{i,j} \in \{0, 1\}, \forall i = 1\ldots g, j = 1\ldots |G_{i}|
\end{aligned}
\end{equation}
where the variable $x_{i,j}$ indicates if item $j$ from group $i$ has been selected or not.

The MCKP is a well-known NP-Hard problem: in this work, we implement a dynamic programming technique to solve it \cite{kellerer2004introduction}. Note that alternative formulations exist, for which a provably optimal greedy algorithm can be constructed: for example, we could consider a fractional formulation of the knapsack problem. This approach, however, would be feasible only if the underlying query execution engine could support partial caching of a relation. As it turns out, the system we target in our work does support hierarchical storage levels for cached relations: what does not fit in RAM, is automatically stored on disk. Although this represents an interesting direction for future work (as it implies a linear time greedy heuristic can be used), in this paper we limit our attention to the 0/1 problem formulation.

\subsection{Query Rewriting}
\label{sec:query_rewriting}
The last step is to transform the original input queries to benefit from the selected combination of \emph{cache plans}.

Recall that the output of a \emph{cache plan} is materialized in RAM after its execution. Then, for each input query that is a \emph{consumer for a given cache plan}, we build an \emph{extraction plan} which manipulates the cached data to produce the output relation, as it would be obtained by the original input query. In other words, in the general case, we apply the original input query to the cached relation instead of using the original input relation. In the case of a \ce{} subsuming identical \se{}s, the extraction plan is an identity: the original query simply replaces the sub-tree containing the \ce{} by its cached intermediate relation. Instead, if shared operators are used -- because of \se{}s having the same fingerprint but different attributes -- we build an extraction plan that applies the original filter and projection predicates or attributes to ``extract'' relevant tuples from the cached relation produced from the \ce{}.

Considering our running example, assume that the output of the MCKP solver is to store $\Omega_2$ and $\Omega_3$ in cache. $\Omega_3$ derives from $\omega_3$, where the composing sub-trees (one from query 1, and one from query 2) are the same, therefore the extraction plan will be $\Omega_3$ itself. Instead, $\omega_2$ (from which $\Omega_2$ derives) contains sub-trees with different filtering and projection predicates: when $\Omega_2$ is materialized in the cache, we need to apply the correct filtering (e.g., ``gender = F'') and projection predicates to extract the actual result when considering the different queries.




%% file: 05_implementation.tex

In this work, we adopt the Apache Spark \cite{zaharia2012resilient} processing system and its extension called SparkSQL \cite{sparksql}.
As anticipated in Section\,\ref{sec:problem}, our prototype implementation glosses over some systems aspects that we discuss next. Queries written in Spark SQL take an abstract form, called a DataFrame. A DataFrame object represents the \emph{logical plan} associated to a query, to produce a given output relation. A logical plan is a tree composed of operators (nodes): thus, each node contains information about the operator type and its attributes (filtering predicates, join columns, etc.).

In Apache Spark, each query issued by a client ``lives'' within an instance of an individual session. The session implements all the machinery necessary to parse and optimize logical plans, build physical plans and schedule low-level tasks that implement the computation required by a given query. As a consequence, each query runs in ``isolation'' and sharing work between queries is thus hindered by the very nature of the Apache Spark architecture.
To enable worksharing across multiple client queries, it is thus necessary to build a centralized component that can accumulate multiple client queries, optimize them, and schedule their execution. We call this component the \emph{SparkSQL Server}. Additionally, it is necessary to modify the typical Apache Spark workflow to submit an application: individual clients should submit their applications to the SparkSQL Server, by passing the logical plans associated to each query.

In this paper, we focus on the implementation of the algorithmic aspects of worksharing in the SparkSQL Server, because our goal is to validate and assess the benefits of the proposed methodology. Hence, we assume the (locally optimized) logical plans associated to each individual client query to be available in the SparkSQL Server, taking the form of a collection (e.g., a list) of DataFrames.
Hence, our prototype implementation materializes as an extension to the existing single query optimizer designed for SparkSQL, namely the Catalyst module \cite{sparksql}. To do so, we follow the optimization process of 4 phases discussed in Sect.\,\ref{sec:caching}. Operator fingerprints are computed to identify all similar subexpressions in the first phase. Phase 2 and phase 4 require query transformations, which we achieve using the Scala's pattern matching and the TreeNode library of Catalyst in SparkSQL. Transforming rules are passed as a function to specify how to transform the logical plan trees. In our prototype, cardinality estimation is achieved by a pre-processing phase, that produces the statistics needed in phase 3. 

Finally, once the worksharing optimization process has produced the caching plans according to the solution to the multiple-choice knapsack problem, the SparkSQL Server behaves as a regular Apache Spark client and submits (sequentially) each rewritten client query to the compute cluster.

%% file: 06_evaluation.tex
We now present experimental results to evaluate the effectiveness of our methodology, which we implement for the Apache Spark and SparkSQL systems.\footnote{Source code of our prototype is available as an open source contribution, available here: \url{https://github.com/DistributedSystemsGroup/spark-sql-worksharing}} First, we focus on a general overview of the performance gains achieved by our MQO approach, using the standard TPC-DS benchmarking; we then proceed with a detailed analysis of caching efficiency for individual operators and simple queries.

\subsection{Experimental setup}
\label{sec:setup}
We run our experiments on a cluster consisting of 8 server-grade worker nodes, with 8 cores each and a 1 Gbps commodity interconnect.

For the macro-benchmark, each worker is granted 30 GB of RAM each, of which half is dedicated to caching. We use the queries in the TPC-DS benchmark library for Spark SQL developed by Databricks~\cite{sparksqlperf}, and generate a CSV dataset with scaling factor of 50.

For the micro-benchmark, each worker is granted 6 GB of RAM each, of which half is used for caching data. The \textbf{synthetic dataset} used for the experiments is stored in HDFS using Parquet and CSV formats. The dataset is a table of 30 columns. The first ten columns $n_i, i=1,2,..10$ are of integer data type, randomly and uniformly generated in the range $[1, 10^{i+2}]$. The next ten columns $d_i, i=1,2,..10$ and the successive ten $s_i, i=1,2,..10$ are of double (in range $[0,1]$) and string (of length 20) data types, respectively. We also vary the input sizes from 10 millions (10M) records (of size 3GB on disk) to 100M records (of size 30GB on disk). Note that, for clarity of exposition, when we display the query plans and input relations for the micro-benchmark we only show a subset of 5 out of the 30 attributes, and label such attributes with names. Hence, for our examples, we call our input relation ``people'', where attribute $n_1$ becomes ``age'', and attribute $s_1$ becomes ``name'', and so forth.

In both benchmarks, we use Apache Spark 2.0. Before running any test, we clear the operating system's buffer cache in all workers and master to obtain more accurate results. We also disable the ``compression on caching data" feature of Spark.

\subsection{Macro-benchmarks}
\label{sec:macro}
We begin with a full-fledged performance benchmark, where we use the standard TPC-DS benchmark adapted to the Apache Spark SQL system \cite{sparksqlperf} to evaluate the benefits of our MQO approach. In particular, we select a subset of all queries available in the TPC-DS benchmark, and focus on the 50 queries that can be successfully executed without failures or parsing errors.

\begin{figure}[t]
	\centering
	\includegraphics[width=0.8\linewidth]{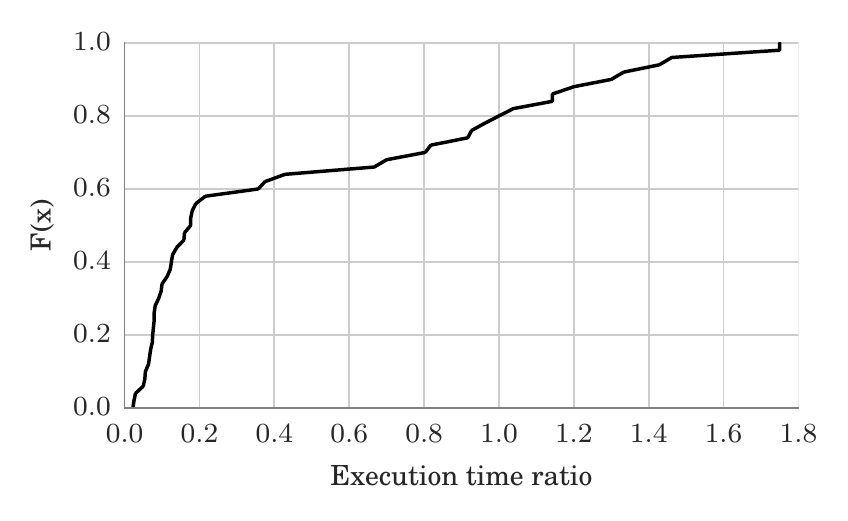}
	\caption{CDF of the performance gains of worksharing for a TCP-DS workload consisting of 50 selected queries.}
	\label{fig:macro-cdf}
\end{figure}

Next, we present results for a setup in which we consider all the 50 queries and execute them in the order of their identifiers, as established by the TPC-DS benchmark. In other words, this experiment identifies all sharing opportunities in a workload consisting of 50 queries, and applies our worksharing optimization to all of them.
Figure\,\ref{fig:macro-cdf} shows the empirical Cumulative Distribution Function (CDF) of the runtime ratios between a system absorbing the workload with MQO enabled and disabled. Overall, we note that, for 60\% of the queries, we obtain a 80\% decrease of the runtime. In total, our approach reduces the runtime for 82\% of the queries. On the other hand, 18\% of the queries experience a larger runtime, which is explained by the overheads associated to caching, as we discuss in Section~\ref{sec:micro}. Overall, our optimizer has identified 60 \se{}s, and it has built 45 \ce{}s. The cache used to store the output of the optimization process is approximately 26 GB (out of 120 GB available). The optimization process took less than 2 seconds, while the query runtime are in the order of tens of minutes (individually) and hours (all together).

Next, we consider an experimental setup in which we emulate the presence of a queuing component that triggers the execution of our worksharing optimization, as anticipated in Sections\,\ref{sec:problem} and \ref{sec:implementation}. In particular, since TPC-DS queries have no associated submission timestamp, we take a randomized approach (without replacement) to select which queries are submitted to the queuing component, and parametrize the latter with the number of queries -- we call this parameter the \emph{window size} -- to accumulate before triggering our MQO mechanism. For a given window size, we repeat the experiment, \ie, we randomly select queries from the full TPC-DS workload, 20 times, and we build the corresponding empirical CDF of the runtime ratio, as defined above. We also measure the number of \se{}s identified within the window size, and show the corresponding empirical CDF. Given this experimental setup, we consider all possible combinations of queries to assess the benefits of worksharing.

\begin{figure}[t]
	\centering
	\includegraphics[width=0.8\linewidth]{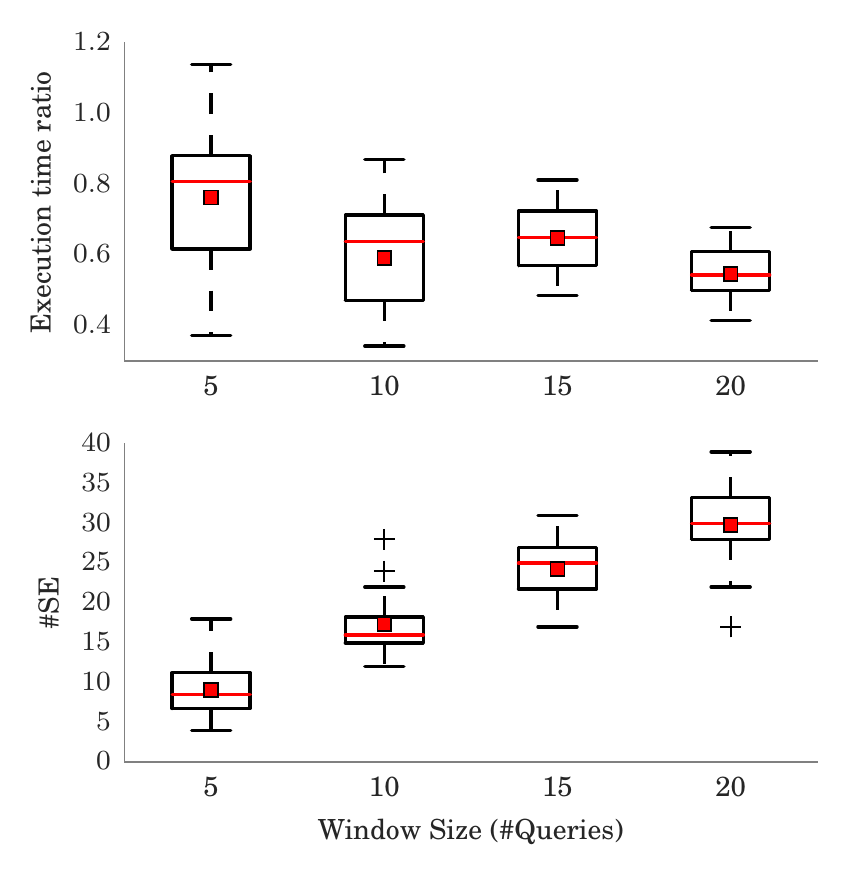}
	\caption{Execution time ratio and number of similar subexpression within a group of queries (given by the size of the window) as the window size increases.}
	\label{fig:macro-workload}
\end{figure}

Figure\,\ref{fig:macro-workload} shows the boxplots of the runtime ratio (top) and number of similar subexpression identified (bottom) for different window sizes. The boxplots indicate the main percentiles (5\%, 25\%, 50\%, 75\%, 95\%) of the empirical CDF, along with the average (red lines).
The Figure shows a clear pattern: as the size of the window increases, there are more chances of finding a high number of \se{}, thus better sharing opportunities, which translates into reduced aggregate runtime. We observe a 20\% decrease of the aggregate runtime (median) with a window size of only five queries, which ramps up to 45\% when the window size is set to 20 queries.

As anticipated in Section\,\ref{sec:problem}, a queuing mechanism can introduce an additional delay for the execution of a query, because the system needs to accumulate a sufficient number of queries in the window before triggering their optimization and execution. Investigating the trade-off between efficiency and delay, as well as studying scheduling policies to steer system behavior is part of our future research agenda.

\subsection{Micro-benchmarks}
\label{sec:micro}
Next, we evaluate our system through a series of experiments based on simple workloads composed by two queries, reading the same synthetic input table, from which we display 5 heterogeneous (numerical and categorical) attributes, for the sake of conciseness. In each experiment, the two queries are run sequentially. We measure and compare the runtime of the Spark jobs associated to each query, according to three strategies: {\it i)} without worksharing, {\it ii)} by having the full input relations cached by the system (FC), {\it iii)} with our worksharing technique (WS). In the latter case, the optimization produces a single \ce{}: hence, the first query triggers the evaluation of the \ce{}, which is cached, whereas the second query benefits the most from worksharing. Each experiment is run three times, and results are averaged.

Overall, our results indicate:
\begin{itemize}
	\item Roughly 50\% and 30\% improvement in \emph{aggregate query latencies} for CSV and Parquet files, respectively.
	\item Our worksharing technique outperforms the naive caching of full input relations.
	\item Between 25\% and 40\% less space used by our method with respect to the full cache technique.
\end{itemize}
In this work we present queries that use \textit{filter} and \textit{project} operators because they appear very frequently in data analysis, where such operators are usually pushed as close as possible to the input tables by traditional query optimization techniques. We also consider the \textit{join} operator that involves shuffling data across the network, but defer discussion to the end of the Section.

\subsubsection*{Filter-based queries}

\begin{figure}[t]
   \centering
   \includegraphics[width=0.99\linewidth]{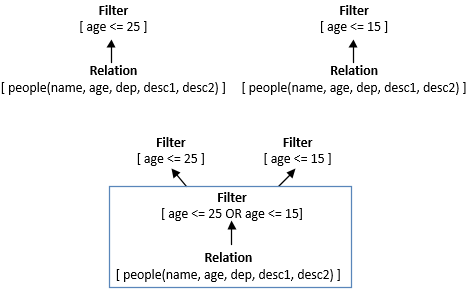}
   \caption{Query (top) and Cache (bottom) plans for Filter-based queries. The input relation is derived from the synthetic dataset, with a labelled schema used for illustration.}
   \label{fig:query1_plans}
\end{figure}
%
%

We consider two simple queries whose logical plan is depicted in Figure\,\ref{fig:query1_plans} (top). Both queries read data from the same input relation, and apply a filter operator with two different predicates on the same attribute. The logical plan displayed at the bottom of the Figure is the output of our multi-query optimizer, that is, a single \ce{} covering both input queries. In this case, the optimized \ce{} can be manually verified: it reads data from the input relation, applies a filter operator with a combined predicate (using the \texttt{OR} logical operator), and caches the intermediate result.\footnote{Note that the \texttt{cache()} operator in Apache Spark is a transformation. As a consequence, it takes effect only upon the first call to an action, with the first (rewritten) query. Thus, the first query effectively ``pays the price'' for caching.} The output of each query can be obtained by applying the filter specific predicate on the cached result.

\begin{figure}[t]
	\centering
   \includegraphics[width=0.9\linewidth]{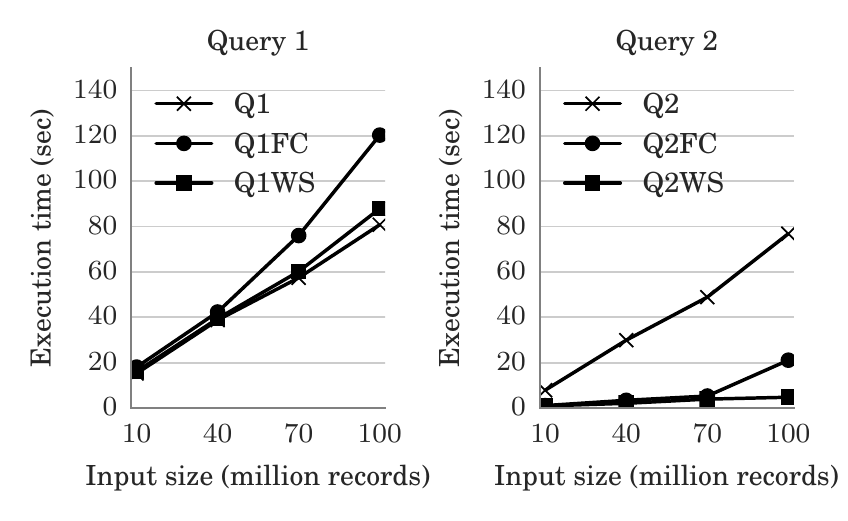}
   \caption{Micro-benchmark for filter-based queries: individual query latencies, comparing baseline, naive caching and our worksharing as a function of input size.}
   \label{fig:query1}
\end{figure}

\begin{figure}[t]
	\centering
	\includegraphics[width=0.9\linewidth]{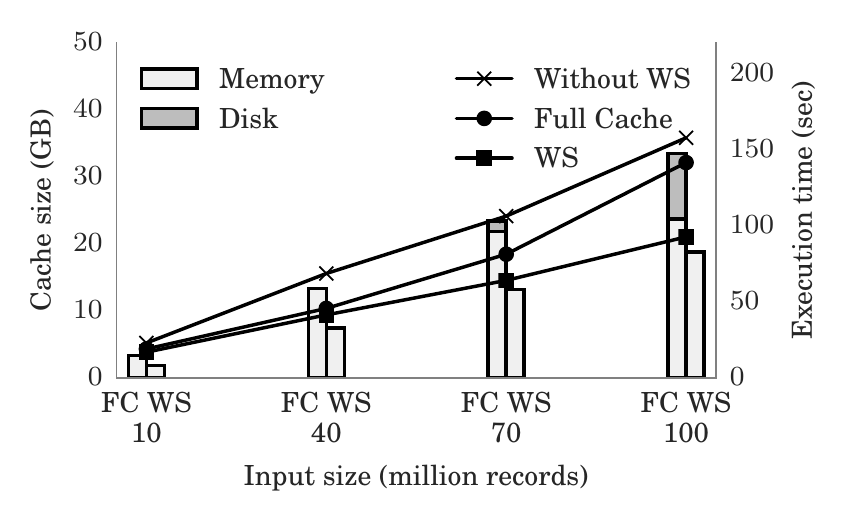}
	\includegraphics[width=0.9\linewidth]{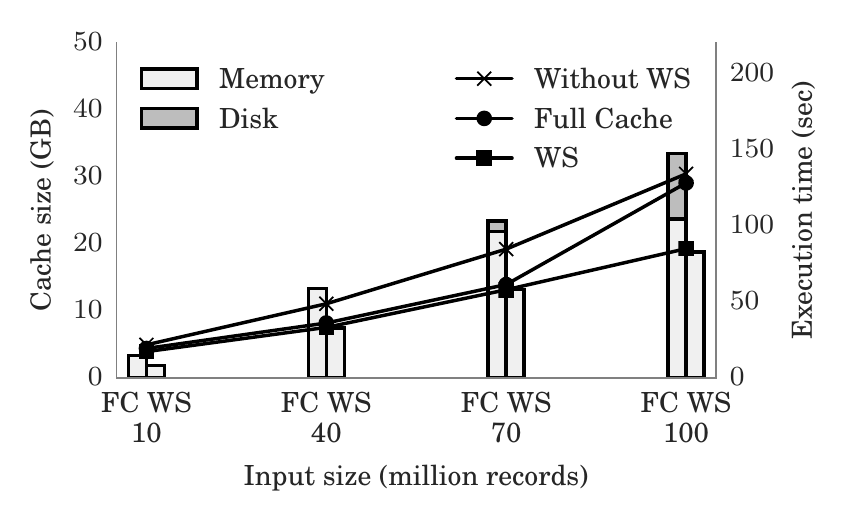}
	\caption{Micro-benchmark for filter-based queries: aggregate query latencies and memory utilization with CSV input format (top) and Parquet input format (bottom).}
	\label{fig:query1summ}
\end{figure}

Next, in Figure\,\ref{fig:query1}, we show the individual query runtime, without any optimization ($Q_i$), using the caching option ($Q_i$FC) and with our optimization ($Q_i$WS) for CSV input files.
In the baseline case, without caching nor worksharing, both queries have similar execution times, that grows linearly with the input size. Instead, when naive caching is used ($Q_i$FC), individual query runtime change: it increases quite dramatically for $Q_1$FC, because this query incurs the overheads associated with a cache operation on the entire input relation, and it decreases for $Q_2$FC, because this query benefits from a cached input. For large input sizes, however, the system spills cached data on disk, with reduced benefits on execution times.
Our MQO approach, instead, achieves superior performance: $Q_1$WS pays a small price for caching just what is needed to produce the output relations, whereas $Q_2$WS runtime becomes negligible and is marginally affected by the input relation size.

The aggregate latency of the workload is shown in Figure\,\ref{fig:query1summ} (for both CSV and Parquet input format). Our worksharing strategy consistently obtain 40\%-50\% aggregate latency improvement with respect to the baseline. Figure\,\ref{fig:query1summ} shows also the system memory utilization dedicated to caching. A naive caching strategy that stores the entire input relation suffers from capacity constraints: once the available RAM is depleted, data is spilled to disk, as visible when the input size increases. Instead, with our worksharing approach, we only cache the output of the combined filtering operation defined by the \ce{}, while satisfying capacity constraints: the RAM is wisely used and spilling to disk is not required.
Overall, the cache size is roughly 25\% of the size of input data size, for both the Parquet and CSV case. Our multi-query optimizer uses only 2/3 of the caching capacity (25\% of 200M records costs 16GB of cache size, which totals 24GB in our system, that is half of 6GB times 8 workers).

\begin{figure}[t]
   \centering
   \includegraphics[width=0.99\linewidth]{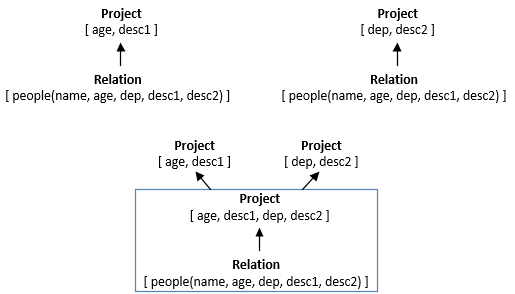}
   \caption{Query (top) and Cache (bottom) plans for Projection-based queries, using synthetic data.}
   \label{fig:query2_plans}
\end{figure}

\subsubsection*{Projection-based queries}
Next, we consider simple queries that only perform projection operations, as shown in Figure\,\ref{fig:query2_plans}. Both queries read data from the same input relation, and apply a project operator on a set of different attributes (top of the Figure). The logical plan displayed at the bottom of the Figure is the optimized \ce{} covering both input queries. The \ce{} can be manually verified: it reads data from the input relation, applies a project operator with the union of each individual query attributes, and caches the intermediate result.

\begin{figure}[t]
	\centering
	\includegraphics[width=0.9\linewidth]{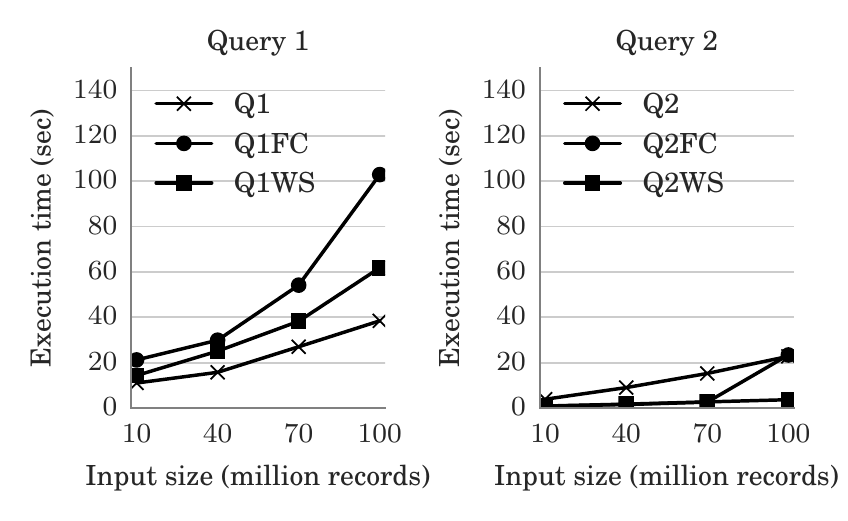}
	\includegraphics[width=0.9\linewidth]{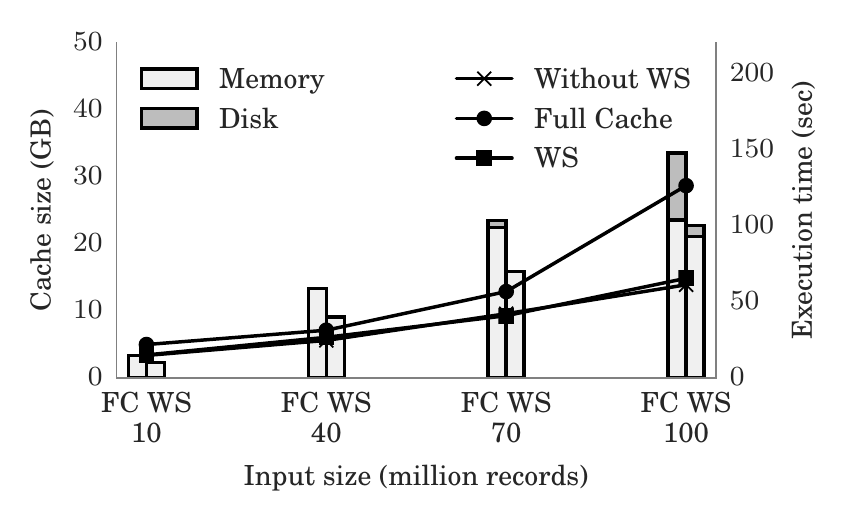}
	\caption{Micro-benchmark for Projection-based queries using Parquet input format: individual query latencies (top), aggregate query latencies and memory utilization (bottom).}
	\label{fig:query2_parquet}
\end{figure}

%
%

Note that projection-based queries can benefit from advanced data representations such as Parquet. In particular, Parquet is geared toward columnar storage: thus, we expect projection queries to execute efficiently, because they only read from disk the data pertaining to selected attributes. It is thus interesting to verify the benefits of caching, if any, for such workloads. In the next series of results, we focus on the Parquet input data types, as shown in Figure\,\ref{fig:query2_parquet}.

Overall, our results share similar qualitative considerations as for filter-based queries: the price to pay to cache data affects the latency of $Q_1$FC and, to a lesser extent, of $Q_1$WS. However, the benefits from caching in $Q_2$FC and $Q_2$WS are less pronounced than for filter-based queries. This is due to the efficiency (for projections) of the Parquet data format. Also, our MQO method is superior in its use of the RAM, when compared to a naive caching strategy.\footnote{The attentive reader might have noticed that also our method eventually spills some contents of the cached data to disk. This is explained by two effects: {\it i)} Apache Spark dynamically adjusts at runtime the amount of memory dedicated to store cached data, and thus overrides the 50\% setting we use in our experiments; {\it ii)} our methodology is based on cardinality estimation to compute the weight of a \ce{}: as a consequence, estimation errors might induce the system to spill some records on disk.}

As shown in Figure\,\ref{fig:query2_parquet} (bottom), the aggregate performance for a simple projection-based workload is in favor of our approach, when compared to naive caching. Nevertheless, we measure no tangible benefit when compared to the baseline: the execution time ``lost'' by query 1 to invest in caching does not pay enough benefits for query 2. Obviously, results obtained with CSV data files are largely in favor of our approach, which outperforms the others in all metrics. Indeed, the parsing costs required for projections translates in non-negligible CPU costs, in addition to increased disk I/O costs due to the requirement to read all data in the input relation.

\subsubsection*{Other types of queries: discussion and summary}
We have conducted an extensive experimental campaign considering queries that combine projection, filtering, as well as join operators. In fact, the generation of synthetic queries to emphasize different aspects related to system performance and the impact of our optimization mechanism, can yield a combinatorial amount of results that we summarize below.

In general, when using the CSV input format, the benefit of worksharing is remarkable: indeed, this wide-spread input data format stresses underlying parallel processing frameworks quite heavily, especially when it comes to parsing, tokenization and type casting costs, as for example discussed in \cite{nsdi15}. Thus, caching achieves consistent gains in query execution time, because it can also save over the ``hidden'' costs related to a specific data format. Conversely, using the Parquet data format can attenuate the benefits of caching, and this is true mainly for simple queries with projections, as discussed above. Instead, when queries involve operators such as joins, for example, other quantities (such as network I/O) are predominant over the savings in data access times, and the effects of caching become tangible again.\footnote{Data compression techniques can be helpful in this case, but we defer their analysis to future work.}

Another way to judge our micro-benchmark results, is to view the query variants we propose as ways to assess different aspects of the cost model defined in this work, and consequently the suitability of the knapsack formulation of the optimization problem. We remark that our results are very robust with respect to the constants we use in our cost model, which is truly desirable as it implies that little to no tuning is required to use our method.


%% file: 07_conclusion.tex
Complex queries for analyzing massive amounts of data have become commonplace today: such trend has been fueled by several efforts to support SQL capabilities on top of large-scale distributed processing frameworks. Similarly to what happens in traditional relational database management systems, users share access to data-intensive processing frameworks and induce workloads with a high degree of redundancy in terms of queries containing similar (sub)expressions.
As a consequence, the traditional problem of \emph{multi-query optimization} that has been largely studied for RDBMSes, also apply to recent data processing frameworks.

In this paper, we presented a new approach to multi-query optimization that uses in-memory caching primitives to improve the efficiency of data-intensive, scalable computing frameworks, such as Apache Spark. Our methodology takes as an input a batch of queries written with the SparkSQL API, and analyzes them to find common (sub)expressions, leading to the construction of an alternative execution plan based on covering expressions, that subsume the individual work required by each query. To make the search problem tractable and efficient, we have used several techniques including: modified hash trees to quickly identify common sub-graphs, and an algorithm to enumerate (and prune) feasible common expressions.
We then cast the multi-query optimization problem as a multiple-choice knapsack problem: each feasible common expression is associated with a value (representative of how much work could be shared among queries) and a weight (representative of the memory pressure imposed by caching the common data), and the goal is to fill a knapsack of a given capacity (representative of memory constraints) optimally.

To quantify the benefit of the proposed methodologies, we implemented a prototype of our method for Apache Spark SQL, and we designed two families of experiments. First, we used the well-known TPC-DS workload to design a macro-benchmark on realistic queries and data. Our results indicated that worksharing opportunities are frequent, and that our proposed methodology brings substantial benefits in terms of reduced query runtime, with up to an 80\% reduction for a large fraction of the submitted queries. Then, using micro-benchmarks, we studied the benefits (and costs) of worksharing for a workload of simple queries, focusing on individual relational operators, including filter-based and projection-based queries. 

In our research agenda, we will consider systems aspects related to the management of a queuing mechanism to accumulate submitted queries in a batch window, and analyze the trade-off that exist between query runtime and execution delay. In this paper, we have obtained promising preliminary results, showing that even small window sizes (that would mitigate execution delays) are sufficient to reap the benefits from worksharing. 

%% file: worsharing.bbl
\begin{thebibliography}{10}

\bibitem{sparksqlperf}
Spark sql performance test.
\newblock \url{https://github.com/databricks/spark-sql-perf}.

\bibitem{agrawal2008scheduling}
P.~Agrawal, D.~Kifer, and C.~Olston.
\newblock Scheduling shared scans of large data files.
\newblock {\em VLDB Endowment}, 1(1):958--969, 2008.

\bibitem{agrawal2000automated}
S.~Agrawal, S.~Chaudhuri, and V.~R. Narasayya.
\newblock Automated selection of materialized views and indexes in sql
  databases.
\newblock In {\em VLDB}, volume 2000, pages 496--505, 2000.

\bibitem{sparksql}
M.~Armbrust{, et al.}
\newblock Spark sql: Relational data processing in spark.
\newblock In {\em Proc. of ACM SIGMOD}, pages 1383--1394, 2015.

\bibitem{arumugam2010datapath}
S.~Arumugam, A.~Dobra, C.~M. Jermaine, N.~Pansare, and L.~Perez.
\newblock The datapath system: A data-centric analytic processing engine for
  large data warehouses.
\newblock In {\em Proc. of ACM SIGMOD}, pages 519--530, 2010.

\bibitem{azim2017recache}
T.~Azim, M.~Karpathiotakis, and A.~Ailamaki.
\newblock Recache: Reactive caching for fast analytics over heterogeneous data.
\newblock {\em VLDB Endowment}, 11(3), 2017.

\bibitem{baril2003selection}
X.~Baril and Z.~Bellahsene.
\newblock Selection of materialized views: A cost-based approach.
\newblock In {\em Advanced Information Systems Engineering}, pages 665--680.
  Springer, 2003.

\bibitem{bhatotia2011incoop}
P.~Bhatotia, A.~Wieder, R.~Rodrigues, U.~A. Acar, and R.~Pasquin.
\newblock Incoop: Mapreduce for incremental computations.
\newblock In {\em Proc. of ACM SoCC}, page~7, 2011.

\bibitem{candea2009scalable}
G.~Candea, N.~Polyzotis, and R.~Vingralek.
\newblock A scalable, predictable join operator for highly concurrent data
  warehouses.
\newblock {\em VLDB Endowment}, 2(1):277--288, Aug. 2009.

\bibitem{candea2011predictable}
G.~Candea, N.~Polyzotis, and R.~Vingralek.
\newblock Predictable performance and high query concurrency for data
  analytics.
\newblock {\em The VLDB Journal}, 20(2):227--248, Apr. 2011.

\bibitem{dalvi2001pipelining}
N.~N. Dalvi, S.~K. Sanghai, R.~Parsan, and S.~Sudarshan.
\newblock Pipelining in multi-query optimization.
\newblock In {\em Proc. of ACM PODS}, pages 59--70, 2001.

\bibitem{dean2008mapreduce}
J.~Dean and S.~Ghemawat.
\newblock Mapreduce: simplified data processing on large clusters.
\newblock {\em Comm. of the ACM}, 51(1):107--113, 2008.

\bibitem{derakhshan2006simulated}
R.~Derakhshan, F.~K. Dehne, O.~Korn, and B.~Stantic.
\newblock Simulated annealing for materialized view selection in data
  warehousing environment.
\newblock In {\em Databases and applications}, pages 89--94, 2006.

\bibitem{dursun2017revisiting}
K.~Dursun, C.~Binnig, U.~Cetintemel, and T.~Kraska.
\newblock Revisiting reuse in main memory database systems.
\newblock In {\em Proc. of ACM SIGMOD}, pages 1275--1289, 2017.

\bibitem{el2015optimization}
A.~El-Helw, V.~Raghavan, M.~A. Soliman, G.~Caragea, Z.~Gu, and M.~Petropoulos.
\newblock Optimization of common table expressions in mpp database systems.
\newblock {\em VLDB End.}, 8(12):1704--1715, 2015.

\bibitem{elghandour2012restore}
I.~Elghandour and A.~Aboulnaga.
\newblock Restore: reusing results of mapreduce jobs.
\newblock {\em VLDB Endowment}, 5(6):586--597, 2012.

\bibitem{finkelstein1982common}
S.~Finkelstein.
\newblock Common expression analysis in database applications.
\newblock In {\em Proc. of ACM SIGMOD}, pages 235--245, 1982.

\bibitem{floratou2016adaptive}
A.~Floratou, N.~Megiddo, N.~Potti, F.~{\"O}zcan, U.~Kale, and
  J.~Schmitz-Hermes.
\newblock Adaptive caching in big sql using the hdfs cache.
\newblock In {\em Proc. of ACM SoCC}, pages 321--333, 2016.

\bibitem{giannikis2012shareddb}
G.~Giannikis, G.~Alonso, and D.~Kossmann.
\newblock Shareddb: Killing one thousand queries with one stone.
\newblock {\em VLDB Endowment}, 5(6):526--537, Feb. 2012.

\bibitem{goldstein2001optimizing}
J.~Goldstein and P.-{\AA}. Larson.
\newblock Optimizing queries using materialized views: a practical, scalable
  solution.
\newblock In {\em ACM SIGMOD Record}, volume~30, pages 331--342, 2001.

\bibitem{gunda2010nectar}
P.~K. Gunda, L.~Ravindranath, C.~A. Thekkath, Y.~Yu, and L.~Zhuang.
\newblock Nectar: Automatic management of data and computation in datacenters.
\newblock In {\em Proc. of OSDI}, volume~10, pages 1--8, 2010.

\bibitem{harizopoulos2005}
S.~Harizopoulos, V.~Shkapenyuk, and A.~Ailamaki.
\newblock Qpipe: A simultaneously pipelined relational query engine.
\newblock In {\em Proc. of ACM SIGMOD}, pages 383--394, 2005.

\bibitem{isard2007dryad}
M.~Isard, M.~Budiu, Y.~Yu, A.~Birrell, and D.~Fetterly.
\newblock Dryad: distributed data-parallel programs from sequential building
  blocks.
\newblock In {\em ACM SIGOPS Op. Systems Review}, volume~41, pages 59--72,
  2007.

\bibitem{ivanova2010architecture}
M.~G. Ivanova, M.~L. Kersten, N.~J. Nes, and R.~A. Gon{\c{c}}alves.
\newblock An architecture for recycling intermediates in a column-store.
\newblock {\em ACM Trans. on Database Systems}, 35(4):24, 2010.

\bibitem{kalnis2002view}
P.~Kalnis, N.~Mamoulis, and D.~Papadias.
\newblock View selection using randomized search.
\newblock {\em Data \& Knowledge Engineering}, 42(1):89--111, 2002.

\bibitem{kellerer2004introduction}
H.~Kellerer, U.~Pferschy, and D.~Pisinger.
\newblock {\em Introduction to NP-Completeness of knapsack problems}.
\newblock Springer, 2004.

\bibitem{li2011platform}
B.~Li, E.~Mazur, Y.~Diao, A.~McGregor, and P.~Shenoy.
\newblock A platform for scalable one-pass analytics using mapreduce.
\newblock In {\em Proc. of ACM SIGMOD}, pages 985--996, 2011.

\bibitem{li2012scalla}
B.~Li, E.~Mazur, Y.~Diao, A.~McGregor, and P.~Shenoy.
\newblock Scalla: a platform for scalable one-pass analytics using mapreduce.
\newblock {\em ACM Trans. on Database Systems}, 37(4):27, 2012.

\bibitem{mtree}
R.~C. Merkle.
\newblock Protocols for public key cryptosystems.
\newblock {\em IEEE Symposium on Security and Privacy}, page 122, 1980.

\bibitem{mistry2001materialized}
H.~Mistry, P.~Roy, S.~Sudarshan, and K.~Ramamritham.
\newblock Materialized view selection and maintenance using multi-query
  optimization.
\newblock In {\em ACM SIGMOD Record}, volume~30, pages 307--318, 2001.

\bibitem{nagel2013recycling}
F.~Nagel, P.~Boncz, and S.~D. Viglas.
\newblock Recycling in pipelined query evaluation.
\newblock In {\em Proc. of IEEE ICDE}, pages 338--349, 2013.

\bibitem{mrshare}
T.~Nykiel, M.~Potamias, C.~Mishra, G.~Kollios, and N.~Koudas.
\newblock Mrshare2: Sharing across multiple queries in mapreduce.
\newblock {\em VLDB Endowment}, 3(1-2):494--505, Sept. 2010.

\bibitem{nsdi15}
K.~Ousterhout, R.~Rasti, S.~Ratnasamy, S.~Shenker, and B.-G. Chun.
\newblock Making sense of performance in data analytics frameworks.
\newblock In {\em Proc. of {USENIX} {NSDI}}, pages 293--307, 2015.

\bibitem{psaroudakis2013sharing}
I.~Psaroudakis, M.~Athanassoulis, and A.~Ailamaki.
\newblock Sharing data and work across concurrent analytical queries.
\newblock {\em VLDB Endowment}, 6(9):637--648, July 2013.

\bibitem{roy2000efficient}
P.~Roy, S.~Seshadri, S.~Sudarshan, and S.~Bhobe.
\newblock Efficient and extensible algorithms for multi query optimization.
\newblock In {\em ACM SIGMOD Record}, volume~29, pages 249--260, 2000.

\bibitem{sellis1988mqo}
T.~K. Sellis.
\newblock Multiple-query optimization.
\newblock {\em ACM Trans. Database Syst.}, 13(1):23--52, Mar. 1988.

\bibitem{shim1999dynamic}
J.~Shim, P.~Scheuermann, and R.~Vingralek.
\newblock Dynamic caching of query results for decision support systems.
\newblock In {\em Proc. of IEEE SSDBM}, pages 254--, 1999.

\bibitem{silva2012exploiting}
Y.~N. Silva, P.-A. Larson, and J.~Zhou.
\newblock Exploiting common subexpressions for cloud query processing.
\newblock In {\em Proc. of IEEE ICDE}, pages 1337--1348, 2012.

\bibitem{sinha1979multiple}
P.~Sinha and A.~A. Zoltners.
\newblock The multiple-choice knapsack problem.
\newblock {\em Operations Research}, 27(3):503--515, 1979.

\bibitem{mqo}
G.~Wang and C.-Y. Chan.
\newblock Multi-query optimization in mapreduce framework.
\newblock {\em VLDB Endowment}, 7(3):145--156, Nov. 2013.

\bibitem{yang1997algorithms}
J.~Yang, K.~Karlapalem, and Q.~Li.
\newblock Algorithms for materialized view design in data warehousing
  environment.
\newblock In {\em VLDB}, volume~97, pages 25--29, 1997.

\bibitem{zaharia2012resilient}
M.~Zaharia, M.~Chowdhury, T.~Das, A.~Dave, J.~Ma, M.~McCauley, M.~J. Franklin,
  S.~Shenker, and I.~Stoica.
\newblock Resilient distributed datasets: A fault-tolerant abstraction for
  in-memory cluster computing.
\newblock In {\em Proc. of USENIX NSDI}, pages 2--2, 2012.

\bibitem{zhang1999genetic}
C.~Zhang and J.~Yang.
\newblock Genetic algorithm for materialized view selection in data warehouse
  environments.
\newblock In {\em DataWarehousing and Knowledge Discovery}, pages 116--125.
  Springer, 1999.

\bibitem{zhou2007efficient}
J.~Zhou, P.-A. Larson, J.-C. Freytag, and W.~Lehner.
\newblock Efficient exploitation of similar subexpressions for query
  processing.
\newblock In {\em Proc. of ACM SIGMOD}, pages 533--544, 2007.

\end{thebibliography}
